\documentclass[10pt,letterpaper,twocolumn]{article}

\usepackage{ol2}
\usepackage{amsmath}
\usepackage{amssymb}

\begin{document}

\twocolumn[

\title{Designing coupled microcavity lasers for high-Q modes with unidirectional light emission}

\author{Jung-Wan Ryu$^*$ and Martina Hentschel}

\address{
Max-Planck-Institut f\"ur Physik komplexer Systeme, N\"othnitzer Str. 38, D-01187 Dresden, Germany
\\
$^*$Corresponding author: jwryu@pks.mpg.de
}

\begin{abstract}
We design coupled optical microcavities and report directional light emission from high-$Q$ modes
for a broad range of refractive indices.
The system consists of a circular cavity that provides a high-$Q$ mode in form of a whispering gallery mode, whereas an adjacent deformed microcavity plays the role of a waveguide or collimator of the light transmitted from the circular cavity. As a result of this very simple, yet robust, concept we obtain high-$Q$ modes with promising directional emission characteristics.
No information about phase space is required, and the proposed scheme can be easily realized in experiments.
\end{abstract}

 ]
Microcavity lasers have attracted much attention because of their high potential in a number of applications related to high-density optoelectric integration \cite{Yam93,Vah03}. An important objective is the design of 
microlasers with unidirectional emission properties. One promising route to achieve this was the modification of the original circular boundary that, due to the inherent rotational symmetry, yields a uniform far field.  
Spiral-shaped microcavity laser \cite{Che03} were one of the first geometries successfully tested in many experiments \cite{Kne04,Ben05,Fuj05,Tul07,Kim08},
and conditions for unidirectional light emission from spiral-shaped microcavities were studied in detail \cite{Aud07,Hen09a,Hen09b}. 
A triangular shape has been proposed especially for small lasers \cite{Kur04,Hen10}. 
In annular microcavities, high-Q modes with unidirectional light emission were achieved through a mechanism that couples a uniform emitting high-$Q$ whispering gallery mode (WGM) to a directional emitting low-$Q$ mode by means of mode hybridization near avoided resonance crossings \cite{Wie06}.
Recently, Lima\c con-shaped microcavity lasers were proposed \cite{Wie08} as source for high-$Q$ modes with unidirectional light emission
working for a substantial range of geometries and material parameters, 
and many experimental demonstrations of unidirectional light emission from those lasers were reported soon after \cite{Yan09,Shi09,Yi09,Son09}. Optimization of the system geometry involved the study of the so-called unstable manifold that governs the chaotic ray dynamics \cite{Sch04,Lee05,Lee07,Shi07}.

Besides the single chaotic microcavity lasers with geometry controlled far-field characteristics, two-disk microcavity lasers have been shown to also produce high-$Q$ modes
with unidirectional light emission \cite{Ryu09}.
These lasers are not practical because there are many nearly degenerate modes with, however, opposite directions of light emission, and moreover, the emission directions depend sensitively on the material parameters such as small variations in the refractive indices.

\begin{figure}[htb]
\centerline{\includegraphics[width=8.3cm]{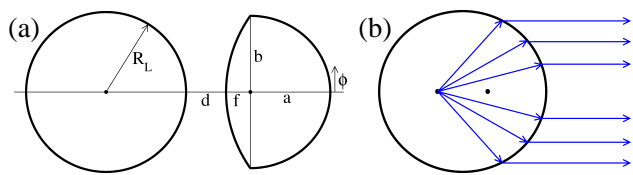}}
\caption{(Color online) (a) Coupled-disk microcavity laser consisting of a circular cavity (left, radius $R_L$) and a circular-elliptic cavity (right, see text for details).
(b) Unidirectional light emissions of light emanating from the focus point of an elliptic cavity with eccentricity being the inverse of its refractive index. 
}
\label{fig1}
\end{figure}

In this Letter we report a scheme that overcomes the drawbacks of two-disk microcavities and the limited range of working refractive indices for Lima\c con-shaped devices, yet keeping their robustness.  Our design couples a circular cavity with a deformed disk, cf.~Fig.~\ref{fig1}. 

Figure 1(a) shows our system which consists of two cavities with different boundary shapes:
a circular cavity of radius $R_L$ (left) and, a distance $d$ apart,
a cavity that is a combination of circular and elliptic shapes, namely 
(i) half of an ellipse with eccentricity $\epsilon$ being inversely proportional to
the refractive index $n$ of the microcavity, $\epsilon=1/n$, (forming the right part of the boundary) and
(ii) a circular arc which passes through and is defined by the three points given by the end points of
the minor axis of the ellipse and, as the third point, its (left) focal point. 
As a result, the boundary shape of the right cavity depends on the refractive index
and, given the semimajor axis $a$ and the refractive index $n$, is defined by the focal length $f=a\epsilon=a/n$
(half the distance between the two focal points)
and the semiminor axis $b=a\sqrt{1-\epsilon^2}=a\sqrt{1-1/n^2}$.
We set $R_L=a=1$ throughout this letter.

In order to explain the motive for the design of the boundary shape in Fig. 1(a), we recall some of the mode properties obtained in previous work on coupled non-identical microcavity lasers \cite{Ryu09,Bor07,Ryu10}.
First of all, modes can be classified into three groups according to their main localization in (i) the left or (ii) the right cavity or (iii) extending over both cavities, the latter were shown to be contained in a ray model  
with deterministic selection rule \cite{Ryu10}. 
Modes in these groups possess different (average) Q-factors.
We are interested in the modes with the highest $Q$-factors as these are the most promising for lasing applications. In our setup, these will be the WGMs living in the left disk, i.e. group-(i) modes, and we use them as the starting point of our study.
As a result, the right cavity plays a same role as a coupler such as a prism or a tapered fiber coupled microsphere laser \cite{Gor94,Cai01}

As shown in Ref. \cite{Ryu09}, optical tunneling into the other (right) disk determines the low-intensity tail structure of these WGMs. It is characterized by two properties.
First, optical tunneling
occurs predominately near the closest approach of the two cavities.
Second, the transmitted light has the same angular momentum profile
as the WGM in the left disk, provided the two disks are sufficiently close.
These two properties provide all the information needed to describe the position and angular momentum profile of the initial light on the right disk, and
we can obtain the resulting far-field pattern using
ray dynamics as described below. Note that the characteristics of the initial light in the second cavity derived in Refs.~\cite{Ryu09,Bor07,Ryu10} for non-identical microdisk laser, is valid in our generalized system because the boundary shape in the area of optical tunneling is circular on either side.

The ray dynamics of the light initiated in the right cavity is then as follows. 
The light originates from the focus of the ellipse, which is part of the circular arc in Fig. 1(a) and reaches the cavity boundary in its elliptic part (other light rays are neglected) where 
the light makes the first dominant transmission through the dielectric interface. Geometric optics for a 
dielectric ellipse with $\epsilon=1/n$ yields \cite{Hec02} that all these rays are 
emitted into the same direction, namely into a far-field angle of $\phi=0$.
This situation is depicted in Fig. 1(b) for a fully elliptic cavity with light emerging from the left focal point. 

\begin{figure}[htb]
\centerline{\includegraphics[width=8.3cm]{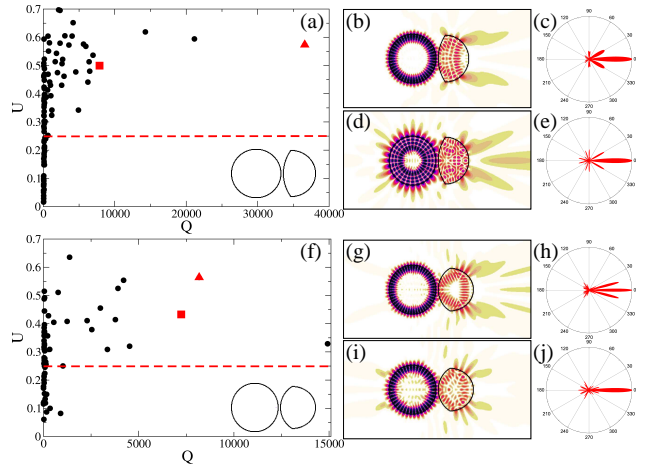}}
\caption{(Color online) $U$-factor as a function of the mode $Q$-factor for $d=0.1$ and (a) $n=3.3$ and (f) $n=2$. 
The red line represents the onset of directional light emission above
$U=0.25$.
The inset shows the system geometry for the respective refractive index $n$. 
Near- and far-field patterns of characteristic high-$Q$ modes, 
are shown in (b),(g) [(d),(i)] and (c), (h)[(e),(j)] for the triangle [square] mode, respectively.
The intensities of the near-field patterns are logarithmically scaled, black and red (dark gray) denote the highest intensities, yellow (light gray) the lowest.}
\label{fig2}
\end{figure}

This geometric optics consideration suggests an intricate interplay between the light in both cavities also in a wave description.
In order to check whether the resulting mode, expected to emit directional into the $\phi=0$ direction, still has the high $Q$-factor of the initial WGM, we investigate the relation between $Q$-factor and unidirectionality. To this end we define the unidirectionality factor $U$ as
\begin{equation}
U = \frac{\int_{-\Phi}^{\Phi} {I(\phi)d\phi}}{\int_{-\pi}^{\pi} {I(\phi)d\phi}} \:,
\end{equation}
where $I(\phi)$ is the light intensity at the far-field angle $\phi$ and we set $\Phi=\pi/4$ in the following, i.e., we consider the ratio between the intensity emitted into a 90 degree window around $\phi=0$ and the total emitted intensity.
The critical value determining unidirectional emission is therefore an $U$-factor larger than $0.25$.
Modes with $U$ sufficiently larger than $0.25$, and  ideally $U$ close to the maximum value 1, possess thus the best unidirectional emission properties.

Figure 2(a) shows the relation between $U$- and $Q$-factor for resonant modes of transverse magnetic (TM) polarization
in the range $20 < \mathrm{Re}(nkR_L) <30$ ($k$ is the wave number in vacuum)
for $n=3.3$ and $d=0.1$, which is obtained numerically using the boundary element method \cite{Wie03}.
Notably, $U$ increases (on average) with $Q$, and we find that a high unidirectional factor $U$ can indeed be combined with a high $Q$-factor, as desired: 
Most of the high-$Q$ modes have $U$-factors $U>0.25$, whereas the $U$-factor of the low-$Q$ modes is distributed around the critical value with a considerable variance. This result proves the relevance of the ray-optics argument developed above.

Examples for high-$Q$, unidirectional emitting modes are indicated by the red triangle and square in Fig.~2(a).
Their near- and far-field patterns, shown in Figs.~\ref{fig2}(b),(d) and (c),(e), respectively, confirm the existence of a WGM localized on the left disk and the unidirectional light emission from the right, deformed cavity.
Although the low-intensity tail structures in the right cavities differ for the two modes -- it is a localized pattern in Fig. 2(b) and a more complex pattern in Fig. 2(d) --,
the directionalities of both far-field patterns are very similar as shown in Figs. 2(c) and (e).

Very similar behavior is found for coupled cavities with refractive index $n=2$, see Fig.~\ref{fig2}(f) to (j). Note that robust unidirectional emission in the proposed scheme is achieved by adjusting the boundary geometry according to the refractive index $n$ [cf.~the insets in Fig.~\ref{fig2}(a) and (f)] and holds, moreover, even for slight deviations in shape and/or $n$. No information about phase space is needed in contrast to single-cavity lasers where the unstable manifold \cite{Sch04,Lee05,Lee07,Shi07} typically provides the basis to design unidirectional emission properties. Consequently, their emission properties strongly depend on the refractive index $n$ - namely on the way the critical line for total internal reflection intersects with the unstable manifold, and for a given geometry, unidirectionality is limited to a certain range of refractive indices. The design introduced here works, however, for a broad range of refractive indices ($n \gtrsim 2.0$) and is therefore applicable to a variety of materials, another useful property adding to the overall robustness of the proposed method. 

\begin{figure}[htb]
\centerline{\includegraphics[width=6.5cm]{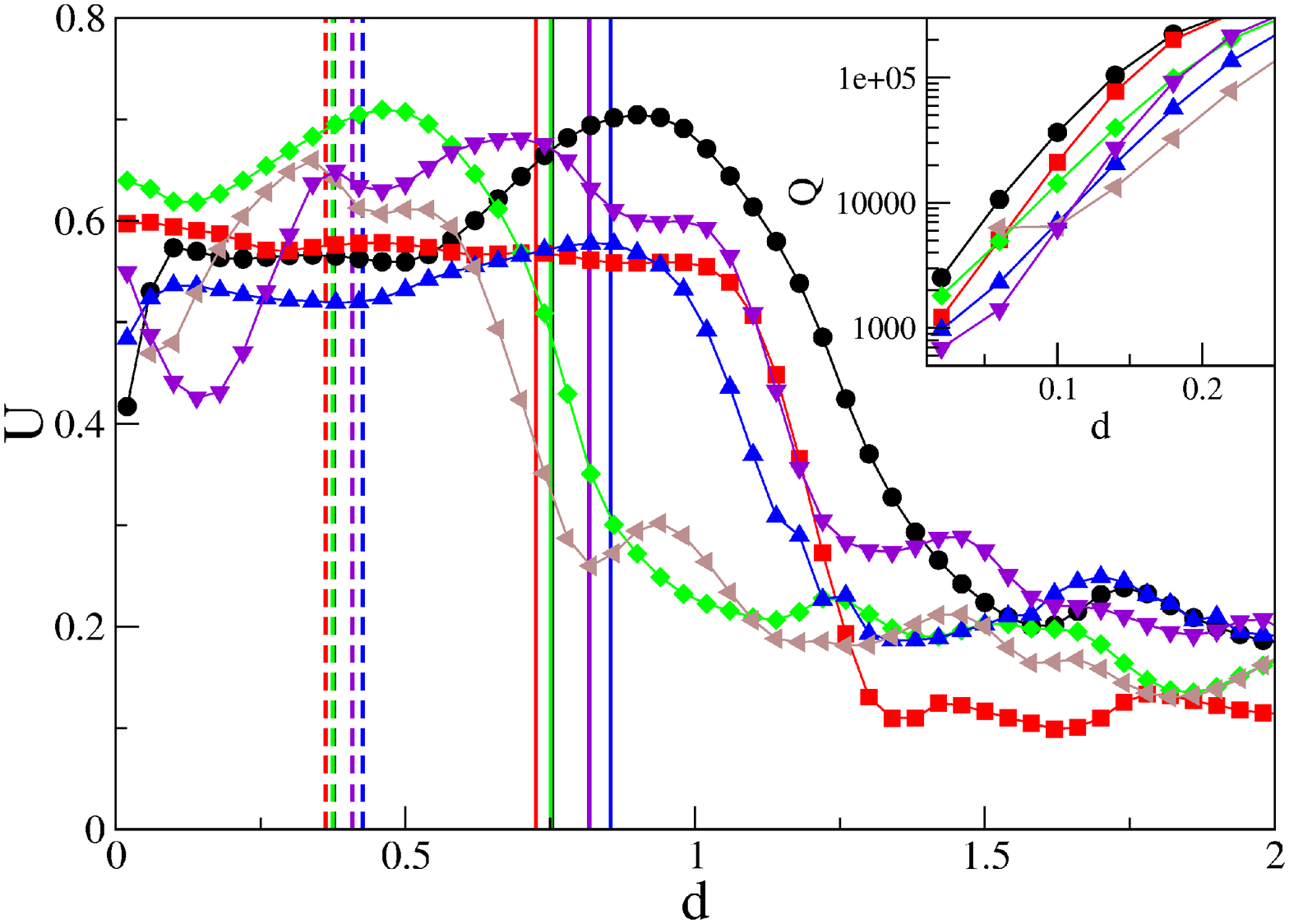}}
\caption{(Color online) $U$-factors of the six modes with the highest $Q$-factors as a function of $d$ (measured in units of $R_L$) for $n=3.3$.
The solid (dashed) lines represent the locations of (half) the vacuum wavelengths of the six modes.
The inset shows $Q$-factors of the six modes as a function of $d$.}
\label{fig3}
\end{figure}

In principle, we can adjust the $U$- and $Q$-factor of a mode by controlling the interdisk distance $d$. This is shown in Fig.~\ref{fig3} for the 
six highest-$Q$ modes as a function of $d$ for $n=3.3$.
For small $d$, especially for $d$ in the vicinity of half the vacuum wavelength, we find good directionality ($U>0.5$), but the $Q$-factor starts to spoil. 
As $d$ increases and exceeds the mode wavelength, the $U$-factor approaches the critical value $0.25$ for isotropic emission, and the $Q$-factor the one of single microdisk WGMs, confirming that optical tunneling as the backbone of the proposed design is lost. 
We point out that $U$ tends to decrease slightly for very small $d$ 
($d$ significantly smaller than half the mode wavelength) due to modifications intunneling \cite{Ryu09}.

We have designed coupled optical microcavities consisting of a circular cavity providing a high-$Q$ WGM that is then tunnel-coupled to a deformed cavity with a $n$-dependent composite geometry (circular arc and half of an ellipse) that guaranties robust unidirectional emission for a broad range of refractive indices from a simple resonator geometry.
No phase space information is needed.
We expect this simple yet efficient concept to be suitable and useful to obtain
high-$Q$ modes with unidirectional emission properties in a number of future applications.


\end{document}